# Wigner time delay of a particle elastically scattered by a cluster of zero-range potentials

M. Ya. Amusia[1, 2], A. S. Baltenkov[3]
and
I. Woiciechowski[4, a]

[1] *Racah Institute of Physics, the Hebrew University, 91904, Jerusalem, Israel*
[2]*Ioffe Physical-Technical Institute, 194021, St. Petersburg, Russian Federation*
[3]*Arifov Institute of Ion-Plasma and Laser Technologies,
100125, Tashkent, Uzbekistan*
[4]*Alderson Broaddus University, 101 College Hill Drive, WV 26416, Philippi, USA*

**Abstract.**
 The Wigner time delay of slow particles in the process of their elastic scattering by complex targets formed by several zero-range potentials is investigated. It is shown that at asymptotically large distances from the target, the Huygens-Fresnel interference pattern formed by spherical waves emitted by each of the potentials is transformed into a system of spherical waves generated by the geometric center of the target. These functions determine flows of particles in and out through the surface of the sphere surrounding the target. The energy derivatives of phase shifts of these functions are the partial Wigner time delay.
 General formulas that connect the s-phase shifts of particle scattering by each of the zero-range potentials with the phases of particle scattering by the potential cluster forming the target are obtained. Model targets consisting of two-, three- and 4-centers are considered. It is assumed that these targets are built from identical delta-potentials with equal distances between their centers.
 The partial Wigner time delay of slow particles by considered targets are obtained. We apply the derived general formulas to consideration of electron scattering by atomic clusters that trap electron near the target, and by calculating the times delay of mesons scattered by few-nucleons systems.

**1. Introduction**
 It is known that the equations of multiple scattering of *s*-waves by a set of fixed in space zero-range potentials is reduced to a system of ordinary algebraic equations [1]. The problem of s-waves scattering by two centers, apparently, was first considered in a paper, devoted to the analysis of the impulse approximation correctness [2]. A detailed analysis of this problem was given by Brueckner in [3] where exact solutions of the wave equation for s-scattering of a particle by two-point scatterers was obtained. The scattering wave function $\psi_\mathbf{k}(\mathbf{r})$ was presented in [3] as a combination of a plane wave and two spherical s-waves generated by scatterers. The amplitudes of these spherical waves are determined by the boundary conditions imposed on $\psi_\mathbf{k}(\mathbf{r})$ at the points where zero-range potentials of the target are located. The asymptotic form $\psi_\mathbf{k}(\mathbf{r} \to \infty)$ determines in closed form the exact elastic scattering amplitude $F(\mathbf{k}, \mathbf{k}')$. Obviously, this method of calculating the scattering amplitude in a closed form is applicable to targets with a larger then two number of delta-centers.

Due to lack of spherical symmetry of the target, the wave function $\psi_\mathbf{k}(\mathbf{r})$ and epy scattering amplitude $F(\mathbf{k},\mathbf{k}')$ cannot be represented as an expansion into a series of spherical functions. However, at asymptotically large distances from the target, where its size can be neglected in comparison with the distance to the observation point, the pair of Huygens-Fresnel spherical waves generated by the target's point potentials transformes into a system of partial spherical waves $\varphi_\lambda(\mathbf{r})$ with the center located at the geometrical center of the target. Phase shifts in the radial parts of these spherical waves $\eta_\lambda(k)$ define the flow of particles inward and outward through the surface of the sphere surrounding the target and also the particle capture time in the considered process.

The time delay and capture time of elastically colliding particles as quantum dynamical observable, were originally introduced in [4-6] by Eisenbud, Wigner, and Smith and was named EWS-time delay. The application of this scattering characteristic ranges from atomic [7-9] to nuclear physics, where it is applied in studies of meson and baryon unstable states [10-12]. The EWS-time delay $\tau(E)$ in particles collision is defined by the energy $E$ derivative of the scattering phase shift $\delta(E)$:

$$\tau(E) = 2\hbar \frac{d\delta(E)}{dE}. \qquad (1)$$

Note that for elastically colliding particles the scattering phases $\delta(E)$ in formula (1) are real.

To calculate the EWS-time delay (1) of a particle colliding with a set of identical zero-range potentials, it is necessary using the function $F(\mathbf{k},\mathbf{k}')$ to establish the relationship between the phase shifts of particle scattering at an isolated delta-potential $\delta_0(E)$ with phases of particle scattering $\eta(E)$ at a multi-center target of delta-potentials. The program of corresponding actions was implemented in [13] where the EWS-time delay of slow electrons colliding with a pair of atoms in the model of non-overlapping atomic spheres was calculated. But in the case of three or more delta-centers in the target, the extraction from the amplitude $F(\mathbf{k},\mathbf{k}')$ the expressions, that connect the phase $\delta_0(E)$ with the phases $\eta(E)$ is rather complicated.

Demkov and Rudakov in [14] developed a partial wave method for a nonspherical scatterer that generalizes the conventional phase method for a spherically symmetric problem. They formulated variational principles that make it possible to calculate above-mentioned partial waves $\varphi_\lambda(\mathbf{r})$ and their phase shifts $\eta_\lambda(k)$ by a direct method, without obtaining the scattering amplitude $F(\mathbf{k},\mathbf{k}')$ in the closed form. In this paper it was shown that for the case when the scatterer can be represented as a set of $N$ zero-range potentials the problem reduces to a purely algebraic one, namely to the inversion of an $N^{th}$ order matrix.

In this article on the base of general formulas developed in [13-15] we will investigate targets that consist of two, three and four zero-range potentials. As far as we know, the EWS-times for such multi-center targets have not been investigated before. The outline of our article is as follows. In Section 2, using a simple example of s-scattering of a wave by a pair of zero-range potentials, we will trace the transformation of the diffraction pattern at large distances from the target. In Section 3, we will derive formulas relating the scattering phases on an isolated delta potential $\delta_0(E)$ to the scattering phases $\eta_\lambda(k)$ upon a pair of such potentials. In Section 4 and 5, the same problem is solved for three- and four-center targets. The derived general formulas are applied to the case of electron scattering by atomic clusters that trap an electron near the target (Section 6), as well to calculation of the EWS-times delay of incoming mesons by few-nucleons systems (Section 7). Section 8 is conclusions.

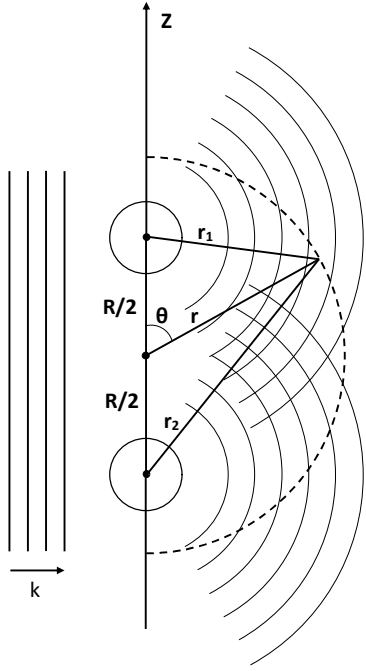

Fig.1. The s-wave scattering by a pair of zero-range potentials separated by a distance $R$; $r$ is an arbitrary observation point.

## 2. Transformation of the diffraction pattern

Consider the s-scattering of a wave by a pair of zero-radius potentials located at a distance $R$ (Fig. 1). Suppose that the scattering phase on each of them is $\delta_0(k)$. The total amplitude of the spherical Huygens waves formed in the process of incident wave scattering upon this axially symmetric target at a point spaced at a distance $r$ from its geometric center is

$$J(\vartheta) \propto \frac{1}{kr_1}\sin(kr_1+\delta_0) + \frac{1}{kr_2}\sin(kr_2+\delta_0). \tag{1}$$

Here the radii $\mathbf{r}_1$ and $\mathbf{r}_2$ are

$$r_1^2 = \frac{R^2}{4} + r^2 - Rr\cos\vartheta;$$

$$r_2^2 = \frac{R^2}{4} + r^2 + Rr\cos\vartheta, \tag{2}$$

where $\vartheta$ is the angle between vectors $\mathbf{R}/2$ and $\mathbf{r}$. The function $J(\theta)$ is shown in Fig. 2. The curves in this figure represent the diffraction pattern profile along a circle with radius $r$. With increasing $r/R$, the amplitude of the alternating function $J(\theta)$ rapidly decreases to zero, that is, the crests of a pair of Huygens waves are transformed into crests of spherical waves generated by the center of the target. The phase shifts of the radial parts of these waves $\eta(k)$ determine the cross section for elastic scattering of a particle by a target formed by a pair of delta-potentials. The formulas connecting the phases $\delta_0(k)$ and $\eta(k)$ are obtained in the next Sections.

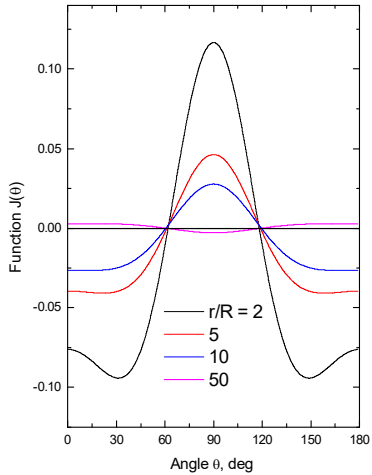

Fig. 2. The function $J(\theta)$ as the diffraction pattern profile along a circle with radius $r$ (dashed curve in Fig. 1).

## 3. Phase shifts $\eta_\lambda(k)$ for particle scattering by a two-center target

According to [14], for the case when the scatterer can be represented as a superposition of $N$ short-range potentials situated at points $\mathbf{R}_j$, the solution of the scattering problem is obtained by imposing boundary conditions, the same as in [3], on the following partial functions

$$\phi_\lambda(\mathbf{r}) = \sum_{j=1}^{N} D_j \frac{\sin(k|\mathbf{r}-\mathbf{R}_j|+\eta_\lambda)}{|\mathbf{r}-\mathbf{R}_j|} \qquad (3)$$

Here $k$ is the particle linear momentum relative to a target. As a result, we obtain the system of homogeneous equations, the solution of which determines the set of phase shifts $\eta_\lambda(k)$. Imposing the following boundary conditions [15] at the localization points of two identical zero-range potentials in the target $\mathbf{R}_j$

$$\phi_\lambda(\mathbf{r})_{\mathbf{r}\to\mathbf{R}_j} \approx C_j\left[\frac{1}{|\mathbf{r}-\mathbf{R}_j|}+k\cot\delta_0\right] \qquad (4)$$

on the wave functions (3) leads to the following homogeneous system of linear equations for the unknown coefficients $D_1$ and $D_2$

$$\begin{aligned} D_1[\operatorname{Im} a\cot\eta+\operatorname{Re} a]+D_2 k(\cot\eta-\cot\delta_0) &= 0 \\ D_1 k(\cot\eta-\cot\delta_0)+D_2[\operatorname{Im} a\cot\eta+\operatorname{Re} a] &= 0 \end{aligned} \qquad (5)$$

Here $a$ is the function $a = \exp(ikR)/R$, where $R$ is the distance between centers. In the system of equations (5) the number of equations is equal to the number of unknowns. Therefore, this system has a nonzero solution if, and only if its determinant equals to zero

$$\begin{Vmatrix} B & A \\ A & B \end{Vmatrix} = (A^2-B^2) = 0. \qquad (6)$$

Here parameters $A$ and $B$ are defined as $A = k(\cos\eta-\sin\eta\cot\delta_0)$ and $B = \sin(kR+\eta)/R$. From (6) we obtain for the case $A+B=0$ the following phase shift for particle-target scattering

$$\cot\eta_0(k) = \frac{kR\cot\delta_0(E)-\cos kR}{kR+\sin kR}. \qquad (7)$$

For the case $A-B=0$ we obtain the second phase shift

$$\cot\eta_1(k) = \frac{kR\cot\delta_0(E)+\cos kR}{kR-\sin kR} \qquad (8)$$

The phases $\eta_\lambda(k)$ in (7) and (8) can be classified by considering their behavior at $k\to 0$ [14, 15]. In this limit the particle wavelength $\lambdabar = 1/k$ is much greater than the target size and the scattering picture should approach that in the case of spherical symmetry. Considering the transition to this limit in (7) and (8), we obtain: $\eta_0(k \to 0) \sim k$ and $\eta_1(k \to 0) \sim k^3$. Thus, the phases (7) and (8) at $k \to 0$ behave similar to the s- and p- phases in the spherically symmetric potential, which explains the choice of their indexes.

Substituting the zero phase $\eta_0(k)$ in (3) we obtain for coefficients at two first wave functions (3) the equality $D_1 - D_2 = 0$. In the limit $r \to \infty$ for partial wave $\phi_0(\mathbf{r})$ in (3) we obtain the following expression

$$\phi_0(\mathbf{r} \to \infty) \propto D_1 \left[ \frac{\sin(k|\mathbf{r}-\mathbf{R}_1|+\eta_0)}{|\mathbf{r}-\mathbf{R}_1|} + \frac{\sin(k|\mathbf{r}-\mathbf{R}_2|+\eta_0)}{|\mathbf{r}-\mathbf{R}_2|} \right]_{r \to \infty}$$
$$= 2D_1 \cos(\mathbf{k}' \cdot \mathbf{R}/2) \left[ \frac{1}{r} \sin(kr+\eta_0) \right]_{r \to \infty}, \quad (9)$$

where $\mathbf{R}_1 = \mathbf{R}/2$ and $\mathbf{R}_2 = -\mathbf{R}/2$; $\mathbf{R}$ defines the position of the target axis in space; vector $\mathbf{k}'$ is particle linear momentum after scattering.

Repeating the same with the phase $\eta_1(k)$ we obtain the equality $D_1 + D_2 = 0$ and for the second partial wave $\phi_1(\mathbf{r})$ in (3) we have the following asymptotic form

$$\phi_1(\mathbf{r} \to \infty) \propto D_1 \left[ \frac{\sin(k|\mathbf{r}-\mathbf{R}_1|+\eta_1)}{|\mathbf{r}-\mathbf{R}_1|} - \frac{\sin(k|\mathbf{r}-\mathbf{R}_2|+\eta_1)}{|\mathbf{r}-\mathbf{R}_2|} \right]_{r \to \infty}$$
$$= -2D_1 \sin(\mathbf{k}' \cdot \mathbf{R}/2) \left[ \frac{1}{r} \sin\left(kr + \frac{\pi}{2} + \eta_1\right) \right]_{r \to \infty} \quad (10)$$

Asymptotic expressions for radial parts of these functions (in squared brackets) coincide with asymptotics of spherical s- and p-waves emitted from geometrical center of target. Whereas, the angle-dependent coefficients at square brackets - functions $Z_0(\mathbf{k}')$ and $Z_1(\mathbf{k}')$ - are analogues of spherical functions, in a series of which the wave function of a particle colliding with a non-spherical target is expanded. For details see [13], where explicit expressions for these functions are obtained. Normalized to the particle unit flow inward and outward through the surface of the sphere surrounding the target, the radial parts of functions (9) and (10), according to [6], determine the EWS time delay of slow particles by the target. In this case, the partial time delays of a particle scattered by a two-center target are determined by formula (1), in which, however, the scattering phase on an individual center is replaced by phases in (9) and (10).

## 4. Phase shifts $\eta_\lambda(k)$ for particle scattered by a three-center target

Imposing the boundary conditions (4) on the wave functions (3) with $N=3$ leads to the following homogeneous system of linear equations for the mixture coefficients $D_1$, $D_2$ and $D_3$

$$D_1 k(\cot\eta - \cot\delta_0) + D_2(\operatorname{Im} a_{12} \cot\eta + \operatorname{Re} a_{12}) + D_3(\operatorname{Im} a_{13} \cot\eta + \operatorname{Re} a_{13}) = 0,$$
$$D_1(\operatorname{Im} a_{21} \cot\eta + \operatorname{Re} a_{21}) + D_2 k(\cot\eta - \cot\delta_0) + D_3(\operatorname{Im} a_{23} \cot\eta + \operatorname{Re} a_{23}) = 0, \quad (11)$$
$$D_1(\operatorname{Im} a_{31} \cot\eta + \operatorname{Re}_{31}) + D_2(\operatorname{Im} a_{32} \cot\eta + \operatorname{Re} a_{32}) + D_3 k(\cot\eta - \cot\delta_0) = 0.$$

The following notations are introduced here: $a_{ij} = \exp(ikR_{ij})/R_{ij}$ and $R_{ij} = |\mathbf{R}_i - \mathbf{R}_j|$. Suppose that the delta-centers in the target are located at the vertices of an equilateral triangle with a side length equal to the distance between centers in two-center target $R = R_{ij}$. The system of equations (11) has a nonzero solution if its determinant equals to zero

$$\begin{Vmatrix} B & A & A \\ A & B & A \\ A & A & B \end{Vmatrix} = (A-B)^2(2A+B) = 0. \tag{12}$$

From equation (11) we obtain three phase shifts of a particle scattered upon the three-center target. The first of them is

$$\cot\eta_0(k) = \frac{kR\cot\delta_0(E) - 2\cos kR}{kR + 2\sin kR} \tag{13}$$

while the second and third ones are

$$\cot\eta_1^{(1,2)}(k) = \frac{kR\cot\delta_0(E) + \cos kR}{kR - \sin kR}. \tag{14}$$

The last two phases (14) coincide with the phase (8) of particle scattering in Section 3.

## 5. Phase shifts $\eta_\lambda(k)$ for a particle scattered by 4-center target

Let us consider the 4-center target in which zero-range potentials are located at the vertices of a tetrahedron with side length equal, as before, to $R$. The model targets discussed in Sections 3-5 exhaust all possible targets in which all center-to-center distances are the same. Imposing the boundary conditions (4) on the wave functions (3) with $N=4$ we obtain, instead of (12), the following equation for the phase shifts

$$\begin{Vmatrix} B & A & A & A \\ A & B & A & A \\ A & A & B & A \\ A & A & A & B \end{Vmatrix} = (A-B)^3(3A+B) = 0. \tag{15}$$

From (15) we obtain four phases of a particle scattered by a 4-center target. The first of them is

$$\cot\eta_0(k) = \frac{kR\cot\delta_0(E) - 3\cos kR}{kR + 3\sin kR}, \tag{16}$$

and the other three ones are

$$\cot\eta_1^{(1,2,3)}(k) = \frac{kR\cot\delta_0(E) + \cos kR}{kR - \sin kR}. \tag{17}$$

It is surprising that in all considered targets only $\eta_0(k)$ phase shifts (7), (13) and (16) differ, while the phases with index $\lambda=1$ are the same. At first glance, this is explained by the equality of the distances between the centers and their identity, that is, by the same boundary conditions imposed on the wave function of the scattering problem. Such an idealization design of considered targets makes it possible to complete the solution of the problem of particle scattering by these nonspherical targets.

In the following sections we will apply above-derived formulas for phase shifts to calculate the cross sections of elastic scattering and EWS-times delay of particle by model targets. The zero-range potential approximation is widely used in describing scattering both in atomic and nuclear physics. If we study the multiply scattering of an electron by a molecule, then the simplest and most natural approximation will be the replacement of each atom in the target by a zero-range potential [15]. The next Section 6 generalizes the results obtained in [13] by considering quantum mechanical scattering of slow electrons by atomic clusters and EWS-time delay in this process.

## 6. Electron elastically scattered by a cluster of atoms

The averaged effective cross section $\bar{\sigma}(k)$ of particle elastically scattered by nonspherical target, that is, the total cross section integrated over all angles of the vector **k'** relative to the vector **k** and then averaged over all angles of vector **k** with respect to the target axes, is determined by the following relation [14]

$$\bar{\sigma}(k) = \frac{4\pi}{k^2} \sum_\lambda \sin^2 \eta_\lambda(k). \tag{18}$$

Here the phase shifts $\eta_\lambda(k)$ are obtained above. The index $\lambda$ by means of which we number the phases for non-spherical targets, in the spherically symmetric case should be replaced by two indices: $l$ and $m$. Summation over magnetic quantum numbers $m$ leads to appearance of the factor $(2l+1)$ in front of the square of the phase sines in (18). Summation here then goes over $l$ only.

For a two-atomic cluster the averaged effective cross section $\bar{\sigma}(k)$ has the following form

$$\bar{\sigma}_2(k) = \frac{4\pi}{k^2}[\sin^2 \eta_0 + \sin^2 \eta_1] = \frac{4\pi}{k^2}[(1+\cot^2 \eta_0)^{-1} + (1+\cot^2 \eta_1)^{-1}]. \tag{19}$$

The phase shifts here are determined by formulas (7) and (8).

For a three-atomic cluster, with phase shifts (13) and (14), the total cross section is equal to

$$\bar{\sigma}_3(k) = \frac{4\pi}{k^2}[\sin^2 \eta_0 + 2\sin^2 \eta_1^{(1,2)}] = \frac{4\pi}{k^2}[(1+\cot^2 \eta_0)^{-1} + 2(1+\cot^2 \eta_1^{(1,2)})^{-1}]. \tag{20}$$

The coefficient 2 in front of the second terms in formula (20) appears due to the fact that two out of three scattering phases, two phase shifts $\eta_1^{(1,2)}(k)$ in (14) coincide.

For 4-atomic target, with phase shifts (16) and (17), the cross section becomes

$$\bar{\sigma}_4(k) = \frac{4\pi}{k^2}[\sin^2 \eta_0 + 3\sin^2 \eta_1^{(1,2)}] = \frac{4\pi}{k^2}[(1+\cot^2 \eta_0)^{-1} + 3(1+\cot^2 \eta_1^{(1,2)})^{-1}]. \quad (21)$$

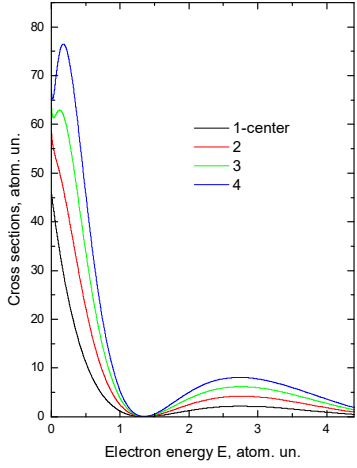

Fig. 3. The averaged effective cross section $\bar{\sigma}(E)$ of particle elastically scattered by targets with 1-4 delta-centers.

The coefficient 3 in front of the second terms in formula (21) appears due to the fact that out of the four scattering phases three phase shifts (17) coincide.

Formulas (20) and (21) generalize the results presented in [13], where the electron scattering cross sections by a cluster with two carbon atoms were calculated. We will assume that inter-atomic distances $R$ in all targets are equal. As concrete value, we chose that in [13], it is equal to the case of the real molecule $C_2$, - $R=2.479$ atomic units (at. un.). In calculations of phase shifts $\eta_\lambda(k)$, as in [13], the phase shift $\delta_0(k)$ is described by the following expression $\delta_0(k) = 2\pi - 1.912k$ [16]. The results of numerical calculations of the electron scattering cross sections by multi-center targets are shown in Fig. 3. There, for comparison, the results of calculating the s-scattering of an electron by an isolated carbon atom are also presented. The curve for a two-center target $\bar{\sigma}_2(E)$ coincides of course with the same curve given in [13], which was calculated both by formula (19) and using the optical theorem [17].

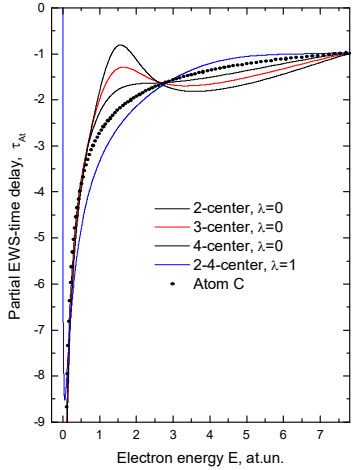

Fig. 4. The partial EWS-time delay $\tau_\lambda(E)$ in electrons collision with targets in units $\tau_{At} = 2{,}419 \cdot 10^{-17}$ s.

The partial EWS-time delay $\tau_\lambda(E)$ in electrons collision with targets is expressed via the derivative in energy $E$ of the scattering phase shift $\eta_\lambda(E)$:

$$\tau_\lambda(E) = 2\tau_{At} \frac{d\eta_\lambda(E)}{dE}. \quad (22)$$

In this formula particle energy $E$ is measured in atomic units and characteristic time is $\tau_{At} = 2{,}419 \cdot 10^{-17}$ s. According to the above formulas, we have three phases of electron scattering with the index $\lambda = 0$. These are given by (7), (13) and (16). And one scattering phase (8), with the index $\lambda = 1$. Numerical results for partial EWS-time delay calculations with these phase shifts are given in Fig. 4. The curves corresponding to $\lambda = 0$ go to $-\infty$ as $E \to 0$, while the curve with the index $\lambda = 1$ vanishes in this limit and rapidly decreases to a value of $\approx -8.5\tau_A$. All curves oscillate around the curve that presents the EWS-time delay by single carbon atom. The curves form a knot at electron energy $E \approx 2.75$ at. un. Analysis shows that in the vicinity of this energy $\eta_\lambda(E) = \pi/2$. Oscillations of curves are associated with the diffraction terms $\sin kR/kR$ and $\cos kR/kR$ in the formulas for the scattering phases. All times delay in Fig. 4 are negative.

This is understandable, since the considered scattering do not have resonance features. A big positive peak can be expected in the vicinity of a resonance. Indeed, let us assume that the phase $\delta_0(E)$ in the single center case has an additional term

$$\delta_{res}(E) \cong \tan^{-1}\left[\frac{\Gamma/2}{E_{res} - E}\right]. \tag{23}$$

It gives a Breit – Wigner addition to EWS-time delay, namely,

$$\Delta\frac{d\delta}{dE} = \frac{\Gamma}{2}\frac{1}{(E_{res} - E)^2 + \Gamma^2/4}. \tag{24}$$

We will meet discuss such a behavior of the scattering phase in the next section.

## 7. Meson elastically scattered by a few-nucleon system

The zero-range potential model is also widely used describing the scattering in nuclear. There this approximation is applied to describe the multiple scattering of mesons by nucleons or the neutrons by nuclei in molecules or crystals [1-3]. In this Section, that develops the results presented in [3], the scattering of slow mesons by two-, three- and four-nucleons systems and respective EWS-time delay in this process will be studied.

The internucleon distances in all considered targets is assumed to be equal to the mean inter-nucleon distance in deuteron $R$ = 2.142 femto-meter (fm) [18]. The scattering phases of a meson by an isolated proton or neutron are assumed to be the same and equal to $\delta_0(E)$. Let us apply formulas (19) - (21) to calculate the cross sections for elastic scattering of mesons by targets of two-, three-, and 4-nucleons in the zero-range potentials model. Assuming, as in [3], that in formulas (7), (8), (13) and (16) the s-scattering of a meson on an isolated nucleon $\delta_0(E)$ is constant and equal to one of three concrete values $\delta_0(E)$ = 20º, 30º, or 45º, we calculate the cross sections $\bar{\sigma}_D(k)$ by formula (19) and $\bar{\sigma}_T(k)$ by formula (20). The ratios of these cross sections to $2\sigma_0(k)$ and $3\sigma_0(k)$, respectively (the meson-nucleon cross section is $\sigma_0 = (4\pi/k^2)\sin^2\delta_0$) are given in Fig. 5. As in the case of deuteron (D) [3], the processes of multiple scattering of mesons on tritium (T) turn out to be decisive in the formation of the total scattering cross section at $\delta_0(k)$ less than 45º. After elementary manipulations, one can show that formulas (19) coincides with the formula (5) in [3], obtained using the optical theorem [17]. Therefore, the curves for deuteron in Fig. 5 coincide with the curves in Fig. 1 in [3]. Numerical calculations of the effective cross sections for 4-nucleon target (21) lead to curves similar to those shown in Fig. 5 for D and T. They are not shown only to make the figure simpler.

The EWS-time delay in the processes of elastic scattering of slow mesons by considered targets is determined by the derivatives of the scattering phases $\eta_\lambda(k)$ with respect to the kinetic energy $E$ of the meson, i.e. by formula similar to (1) and (22), but with phases $\eta_\lambda$ instead of $\delta_l$ and $\tau_{Nuc}$ instead of $\tau_{At}$:

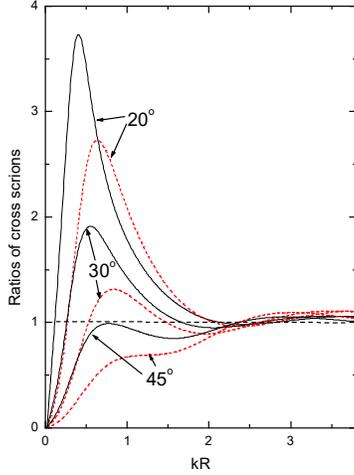

$$\tau_\lambda(E) = 2\tau_{Nuc} \frac{d\eta_\lambda(E)}{dE}, \quad (25)$$

where the meson energy $E$ is measured in MeV and the characteristic time is $\tau_{Nuc} = \hbar/\text{MeV} = 6.582 \cdot 10^{-22}$ s. When calculating the curves in Fig. 5, following [3], we assumed the phase shifts for s-scattering on a single nucleon $\delta_0(E)$ to be constant.

Fig. 5. The ratios of cross sections $\bar\sigma_D(k)/2\sigma_0(k)$ - solid lines calculated with formula (19) and $\bar\sigma_T(k)/3\sigma_0(k)$ - dashed lines calculated with formula (20). The meson-nucleon cross section is $\sigma_0 = (4\pi/k^2)\sin^2\delta_0$.

When calculating $\tau_\lambda(k)$ using formula (25), we need also associate a specific meson wavelength $1/k$ with the various values of the phase shift $\delta_0(E)$. Numerous experimental data for phase shifts as functions of meson linear momentum $k$ are available in the literature in the energy range $E$ between tens and hundreds MeV. We have chosen the empirical dependence obtained in [19-21]. The nuclear phase shifts $\delta_0(E)$ are fitted with an analytical function, which incorporates the threshold behavior (three first terms in formula below) and a term, which represents the nearest π-nucleon resonance:

$$\frac{\tan\delta_0(E)}{q} = b + fq^2 + dq^4 + \frac{x\Gamma_0\omega_0 q_0^{-1}}{\omega_0^2 - \omega^2}, \quad (26)$$

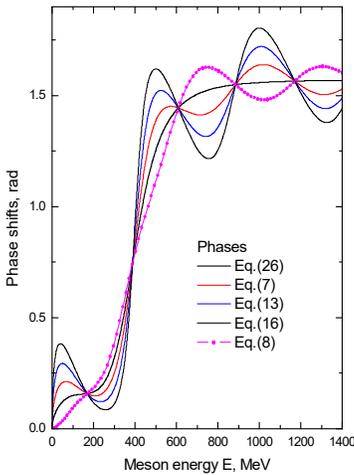

Fig. 6. The scattering phases of mesons on all considered targets.

where $q$ is the center-of-mass linear momentum and $\omega$ is the center-of-mass-energy of π-nucleon collision. Recall that, as in Brueckner's article [3], we consider the target to be infinitely heavy. The wave number $q = c\hbar k$ in (26) is expressed in units of MeV/$c$; $k$ is the meson wave number in inverse fm. The constants $b, f,$ and $d$ measured in the corresponding powers of MeV/c were obtained in [21] (Table II, the first row) to achieve the best fit of the expression (21) with the large set of experimental data. We also used for the fixed resonance parameters $x$, $\omega_0$, $q_0$ and $\Gamma_0$ the constants, given in first row in Table 1 of paper [19].

The results of calculating the phase (26) and the scattering phases of mesons on all considered targets are presented in Fig. 6. The scattering phases $\delta_0(E)$ represent monotonically increasing smooth curve. Whereas the scattering phases $\eta_\lambda(E)$ on the considered targets oscillate around this curve. Moreover, the curves for phases with indices λ=0 and λ=1 oscillate in antiphase. The appearance of these oscillations in the phases is associated with the diffraction of meson waves on scattering centers of the targets.

In the formulas for phase shifts $\eta_\lambda(k)$ obtained above, the latter are represented as functions of the meson momentum $k$, which is related to its kinetic energy $E$ by the following relation [22]

$$k = \frac{\sqrt{E(E+2m_0c^2)}}{\hbar c} = \frac{1}{\hat{\lambda}}, \qquad (27)$$

where $c\hbar = 197.326$ MeV·fm and for π-meson the rest mass is $m_0 = 139.57$ МэВ/$c^2$ [22].

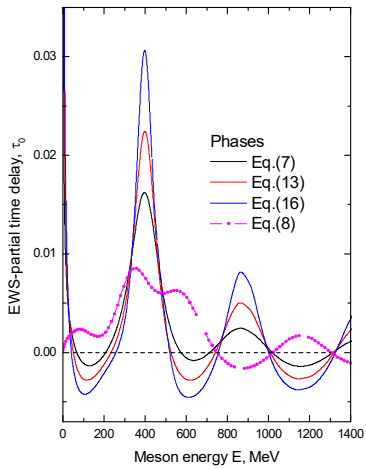

Fig. 7. Partial EWS-times delay, as functions of the meson energy $E$, in units $\tau_{Nuc} = \hbar/\text{MeV} = 6.582 \cdot 10^{-22}$ s.

The results of calculating the partial EWS-times delay of mesons are shown in Fig. 7. Partial EWS-times delay, as functions of the meson energy $E$, in these figures are oscillating curves. Most part of the curves is located in the positive half-plane of the figure. The main peak of the curves falls on meson energy $E$ of the order of 400 MeV. The curve corresponding to the energy derivative of the phase shift with index λ=1, determined by formula (8), turns out to be in antiphase with the rest of the curves at $E > 600$ MeV.

## 8. Conclusions

In the present work, we discuss an original method for calculating phase shifts and partial EWS-time delay of particles elastically scattered by targets that consist of only a few zero-range potentials. A multicenter target does not have spherical symmetry. Therefore, the wave function of a scattering particle cannot be represented as an expansion into a series of spherical functions. However, at asymptotically great distances from the target, the latter can be considered as a point source of spherical waves, the phase shifts of which determine the scattering cross section of a non-spherical target. Wave functions (3) are used to calculate the scattering phases by a target. Imposing boundary conditions on these functions leads to equations for the phases. The existence of variational principles, as shown in [14, 15], makes it possible to calculate the scattering phases on a system of zero-range potentials by a direct and simple method. By analogy with the spherically symmetric case, there is every reason to believe that the partial wave method is the most convenient for studying and calculating particle scattering by nonspherical systems. Such an approach to the calculation of scattering phases on a system of centers to the best of our knowledge has not been discussed in the literature earlier.

The group of targets considered here is the maximal idealized model constructions that make it possible to complete the solution of the particle scattering problem. The scattering phases of a particle by all zero-range potentials are considered, as in [3, 13], to be the same. The targets at the moment of a collision are assumed to be motionless. The distances between centers in all targets are assumed to be equal to *R*. Note, by the way, that the considered group exhaust all possible targets in which all the distances between the scattering centers are the same. Employed approximations maximally simplify the form of determinants (6), (12) and (15), and therefore the calculation of the scattering phases becomes easy.

It is obvious that the replacement of identical phases $\delta_0(E)$ by different ones, or introduction of different distances *R* between scattering centers, transforms the determinants, which complicates the calculation of the phases, however, does not make their calculation impossible. The cross sections of elastic scattering and the EWS times delay as functions of inter-centers distances should be averaged, over the system wave functions. However, we did not do it, since our goal was to estimate in the first approximation the magnitude of the EWS-time delay for target systems with a few identical centers. It turned out that our times for mesons are of the same order of magnitude as times considered, for example, in articles [5, 16]. The existence of positive time delay peaks in Fig. 7 is one of the main observations of the present work, because a positive maximum in time delay is a necessary condition for the existence of real resonances. Their appearance in complex targets, as shown in this paper, is associated with a resonance in the cross section of meson-nucleon collision. Their experimental search and study are, in our opinion, interesting problems.

## References

1. M. L. Goldberger and K. M. Watson, *Collision Theory* (New York: Wiley, 1964).
2. G. F. Chew and G. C. Wick, Phys. Rev. **85**, 636 (1952).
3. K. A. Brueckner, Phys. Rev. **89**, 834 (1953).
4. L. E. Eisenbud, Ph. D. thesis. Princeton University (1948).
5. E. P. Wigner, Phys. Rev. **98**, 145 (1955).
6. F. T. Smith, Phys. Rev. **118**, 349 (1960).
7. U. Fano and A. R. P. Rau, *Atomic Collisions and Spectra* (Academic, Orlando, 1986).
8. N. Yamanaka, Y. Kino, and A. Ichimura, Phys. Rev. A **70**, 062701 (2004).
9. Chun-Woo Lee, Phys. Rev. A **58**, 4581 (1998).
10. J. L. Agudín, Phys. Rev. **171**, 1385 (1968).
11. N. G. Kelkar, M. Nowakowski and K. P. Khemchandani, J. Phys. G: Nucl. Part. Phys. **29**, 1001 (2003).
12. P. Danielewicz and S. Pratt, Phys. Rev. C **53**, 249 (1996).
13. M. Ya. Amusia and A. S. Baltenkov, JETP **131**, 707 (2020).
14. Yu. N. Demkov and V. S. Rudakov, Soviet Phys. JETP, **32**, 1103 (1971).
15. Yu. N. Demkov, V. N. Ostrovskii, *Zero-Range Potentials and Their Applications in Atomic Physics* (Boston: Springer Publ. 1988).
16. A. S. Baltenkov and A. Z. Msezane, Eur. Phys. J. D **71**, 305 (2017).
17. L. D. Landau and E. M. Lifshitz, *Quantum Mechanics, Non-Relativistic Theory* (Pergamon Press, Oxford, 1965).
18. R. Pohl *et al*, Metrologia, **54**, L1 (2017).
19. G. Rowe, M. Salomon, and R. H. Landau, Phys. Rev. C **18**, 584 (1978).
20. R. A. Arndt, I. I. Strakovsky, R. L. Workman, and M. M. Pavan, Phys. Rev. C **52**, 2120 (1995).
21. A. A. Ebrahim and R. J. Peterson, Phys. Rev. C **54**, 2499 (1996).
22. H. A. Bethe and F. Hoffmann, *Meson and Fields, volume II, Mesons* (ROW, Peterson and Company, New York, 1955).
23. N. G. Kelkar, J. Phys. G: Nucl. Part. Phys. **29**, L1 (2003).

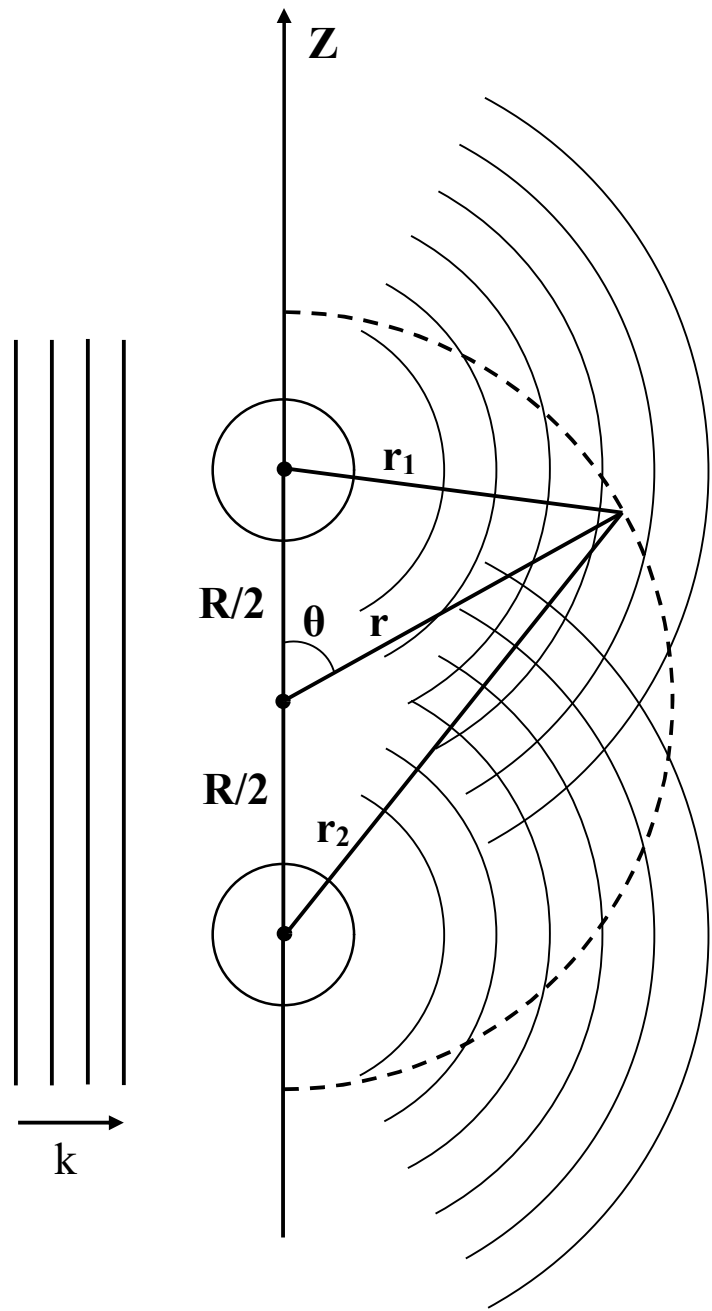

Fig.1. The s-wave scattering by a pair of zero-range potentials separated by a distance $R$; $r$ is an arbitrary observation point.

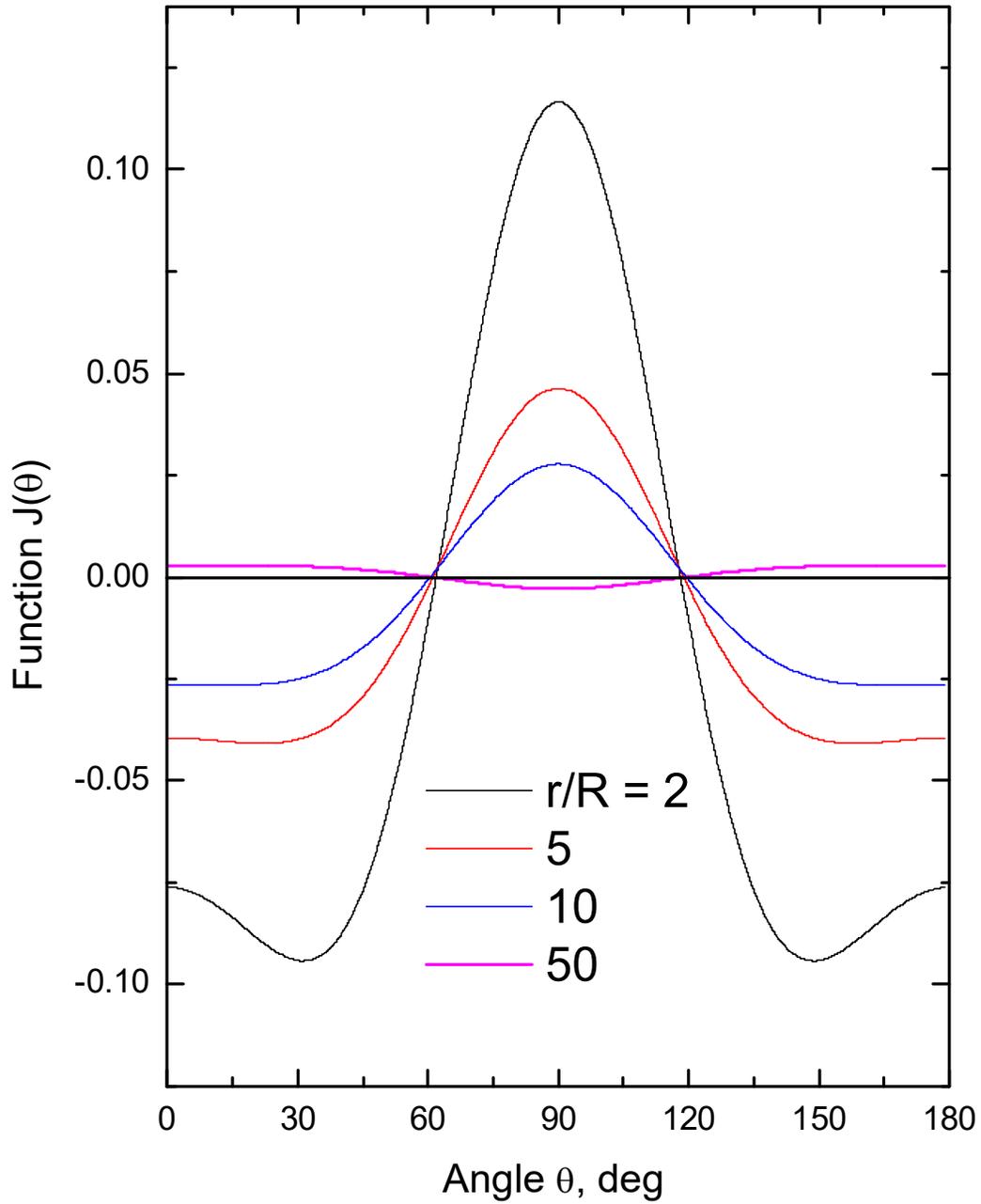

Fig. 2. The function $J(\theta)$ as the diffraction pattern profile along a circle with radius $r$ (dashed curve in Fig. 1).

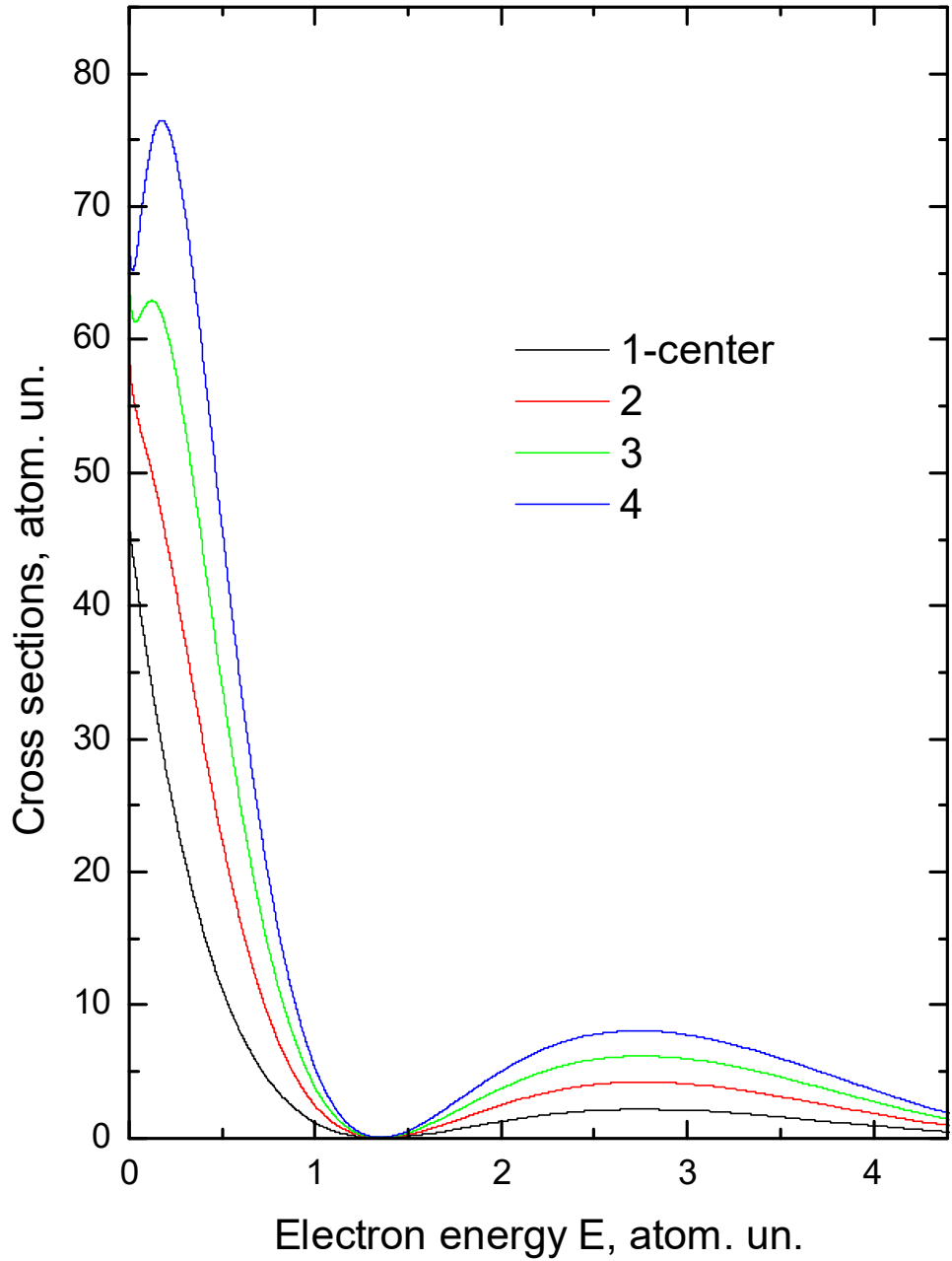

Fig. 3. The averaged effective cross section $\bar{\sigma}(E)$ of particle elastically scattered by targets with 1-4 delta-centers.

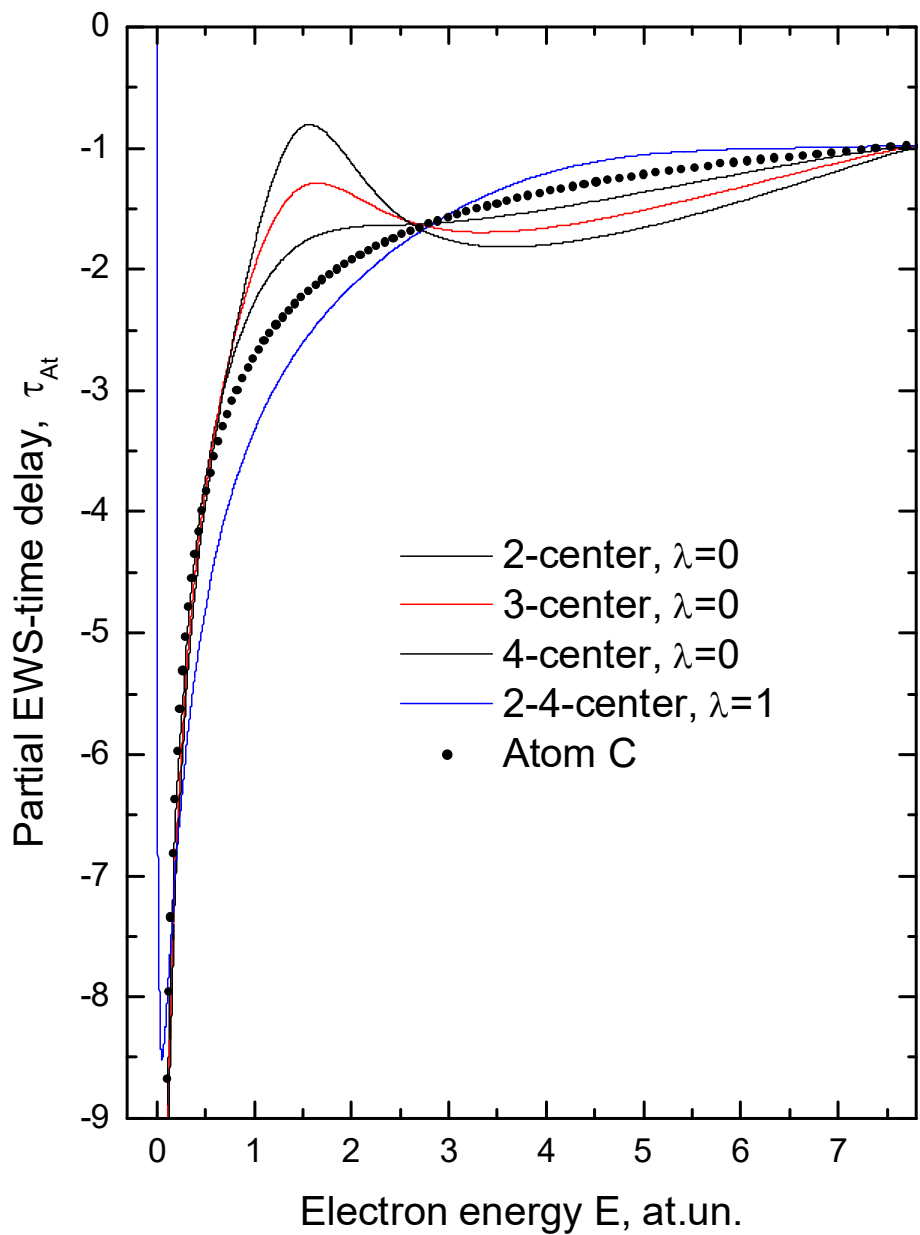

Fig. 4. The partial EWS-time delay $\tau_\lambda(E)$ in electrons collision with targets, in units $\tau_{At} = 2{,}419 \cdot 10^{-17}$ s.

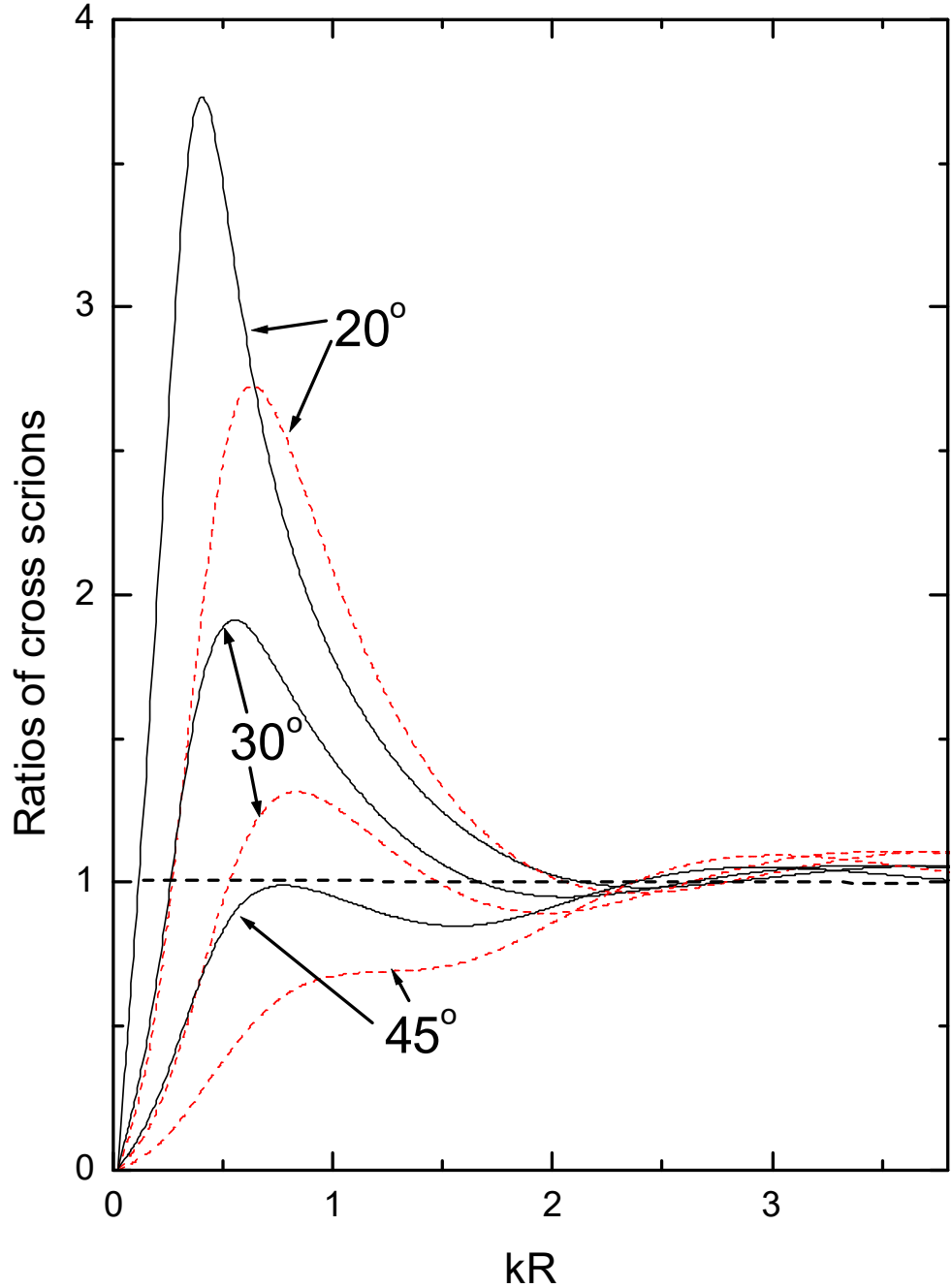

Fig. 5. The ratios of cross sections $\bar{\sigma}_D(k)/2\sigma_0(k)$ - solid lines calculated with formula (19) and $\bar{\sigma}_T(k)/3\sigma_0(k)$ - dashed lines calculated with formula (20). The meson-nucleon cross section is $\sigma_0 = (4\pi/k^2)\sin^2\delta_0$.

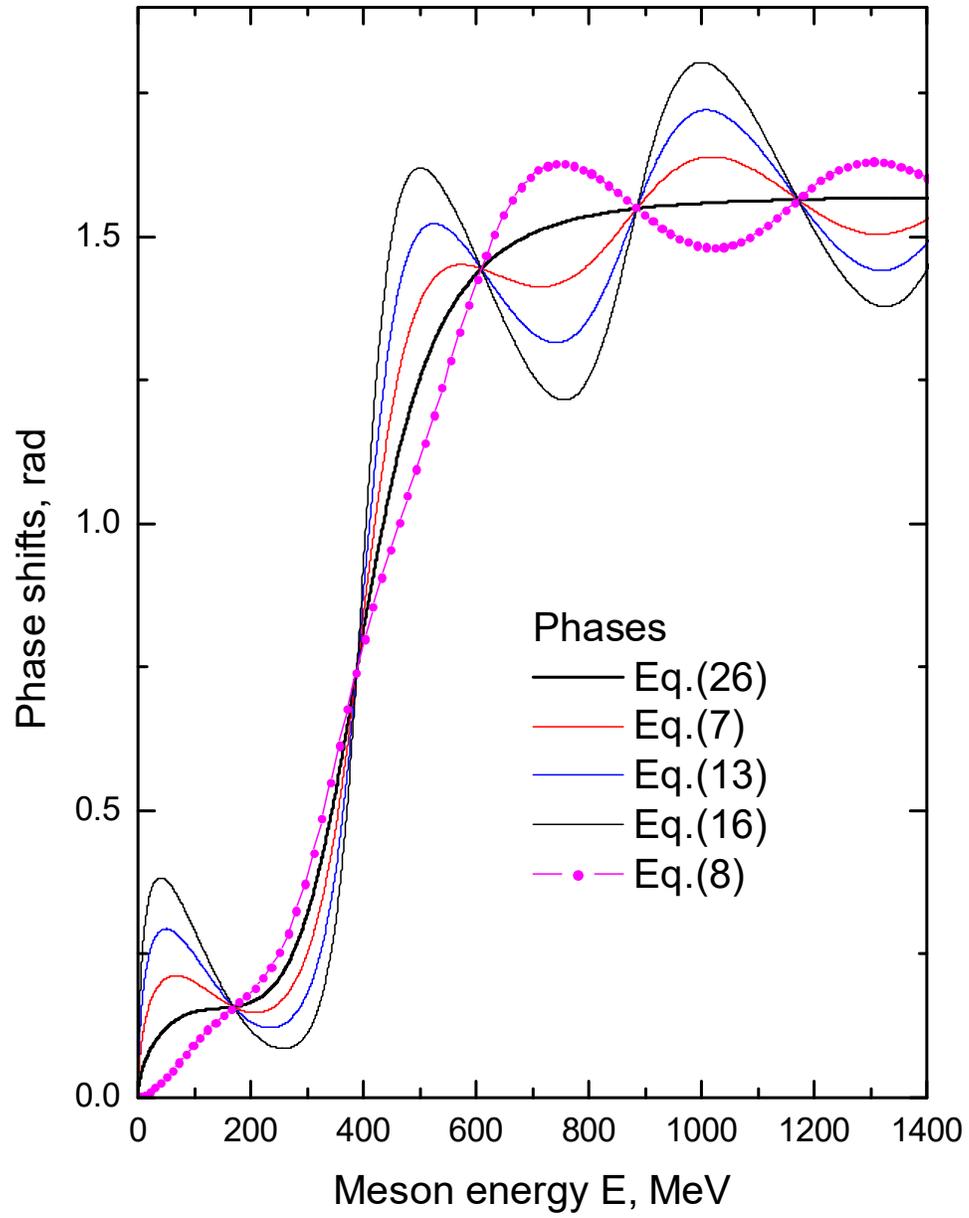

Fig. 6. The scattering phases of mesons on all considered targets.

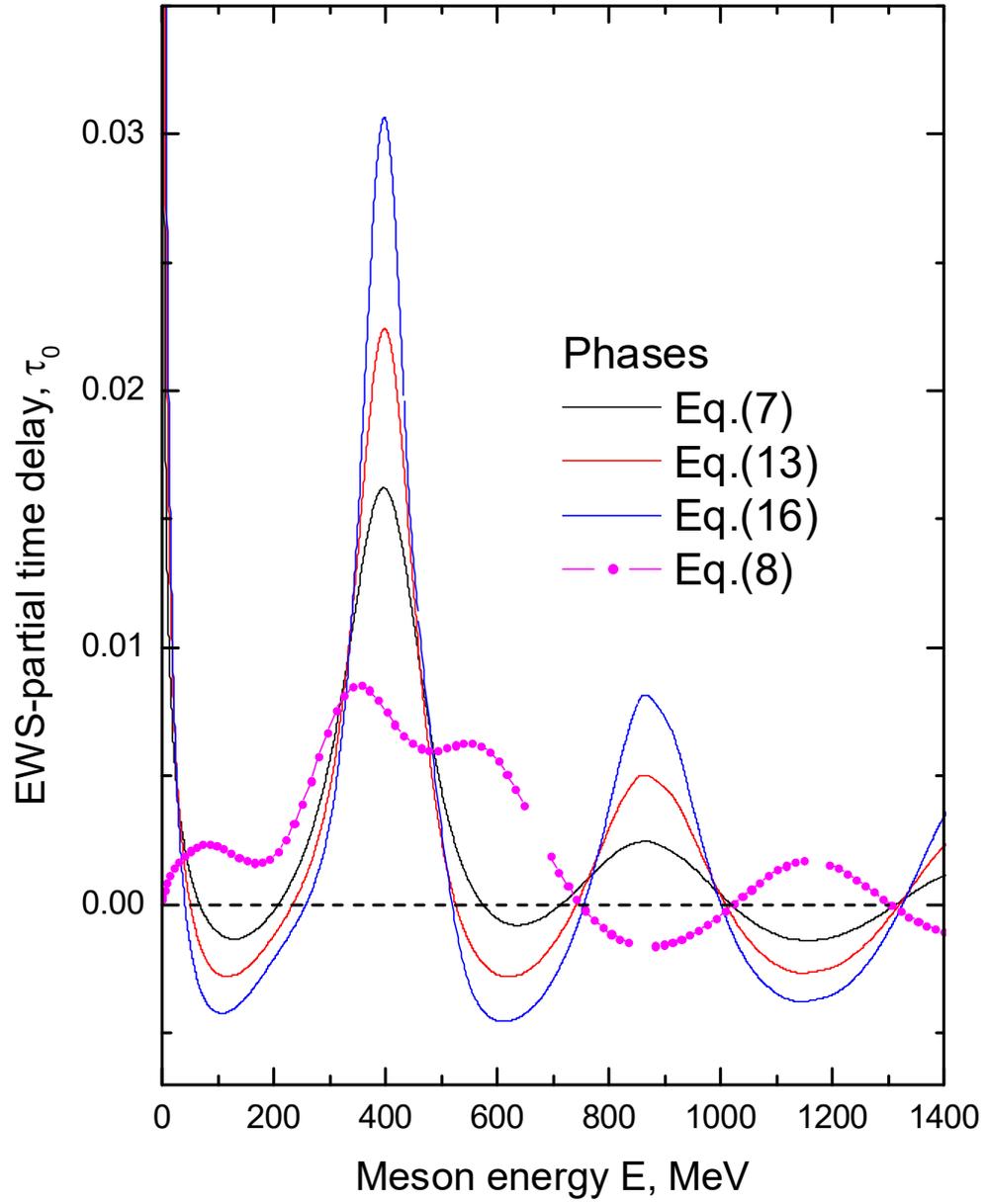

Fig. 7. Partial EWS-times delay, as functions of the meson energy $E$, in units $\tau_{Nuc} = \hbar/\text{MeV} = 6.582 \cdot 10^{-22}$ s.